\documentclass[a4paper,11pt]{article}
\usepackage{pos}
\usepackage{cleveref}
\crefname{figure}{Fig.}{Figs.}

\title{The Milky Way is a Laboratory for New Ultra-long-baseline Neutrino Physics}

\author*[a]{Miller MacDonald}
\author[a]{Kiara Carloni}
\author[a]{Carlos A. Argüelles}
\author[c]{Rafael Alves Batista}
\author[b]{Ivan Martínez-Soler}

\affiliation[a]{Harvard University, Department of Physics and Laboratory for Particle Physics and Cosmology,\\
  Cambridge, MA 02138, USA}

\affiliation[b]{Institute for Particle Physics Phenomenology,
Durham University,\\ South Road, DH1 3LE, Durham, UK}

\affiliation[c]{Sorbonne Université, CNRS, UMR 7095, Institut d’Astrophysique de Paris,\\ 98 bis bd Arago, 75014 Paris, France}

\emailAdd{mmacdonald@college.harvard.edu}
\emailAdd{kcarloni@g.harvard.edu}
\emailAdd{carguelles@fas.harvard.edu}
\emailAdd{ivan.j.martinez-soler@durham.ac.uk}
\emailAdd{rafael.alves\_batista@iap.fr}

\abstract{The IceCube Neutrino Observatory recently published evidence for diffuse neutrino emission from the Galactic Plane at $4.5\sigma$ significance. This new source of astrophysical neutrinos provides an exciting laboratory for probing the nature of neutrino masses. In particular, extremely small mass splittings, such as those predicted by quasi-Dirac neutrino mass models, and finite neutrino lifetimes from neutrino decays, would induce effects on the spectra and flavor ratios of neutrinos with TeV-scale energies traversing kiloparsec-scale baselines. Using \texttt{TANDEM}, an upcoming three dimensional galactic neutrino emission model, we explore the sensitivity of IceCube and KM3NeT/ARCA to these ultra-long-baseline phenomena. We find that a combined analysis would be sensitive to quasi-Dirac mass splittings $10^{-14.0}~\mathrm{eV^2} \lesssim \delta m^2 \lesssim 10^{11.6}~\mathrm{eV^2}$ and neutrino lifetimes $m / \tau \gtrsim 10^{-14.1}~\mathrm{eV^2}$ at $> 1\sigma$, both regions constituting as-yet unexplored parameter space. Our results demonstrate the potential that astrophysical neutrino sources and global neutrino telescope networks have in probing new regions of exotic neutrino mass models.}

\ConferenceLogo{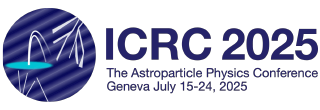}

\FullConference{39th International Cosmic Ray Conference (ICRC2025)\\
 15–24 July 2025\\
Geneva, Switzerland\\}

\begin{document}
\maketitle

\section{Introduction}

Neutrino oscillations were discovered around the turn of the century \cite{Super-Kamiokande:1998kpq, SNO:2002tuh}, implying that at least two neutrino mass states have nonzero mass and providing one of the strongest pieces of evidence for the existence of physics Beyond the Standard Model (BSM). Neutrino masses lead to two classes of BSM models; 1) the neutrino mass mechanism, still unknown, and 2) novel phenomena arising from nonzero neutrino mass states.

Many exotic neutrino models extending the Standard Model (SM) lead to modifications to neutrino vacuum propagation over ultra-long baselines. These effects scale like $L/E$, where $L$ is the neutrino propagation distance and $E$ is the neutrino energy. Examples of such models fall into both of the referenced classes above. For example, the quasi-Dirac (QD) neutrino model \cite{deGouvea:2009fp} postulates the existence of very small Majorana terms in the neutrino mass matrix, leading to maximally mixed new sterile neutrino states that form near-degenerate active-sterile pairs with the three active mass states. Additionally, massive neutrinos present the possibility of neutrino decay, either into undetectable products (``invisible'' decay) or active lighter mass states (``visible'' decay) \cite{Beacom:2002cb}. 

\begin{figure}
\centering 
\includegraphics[width=\linewidth]{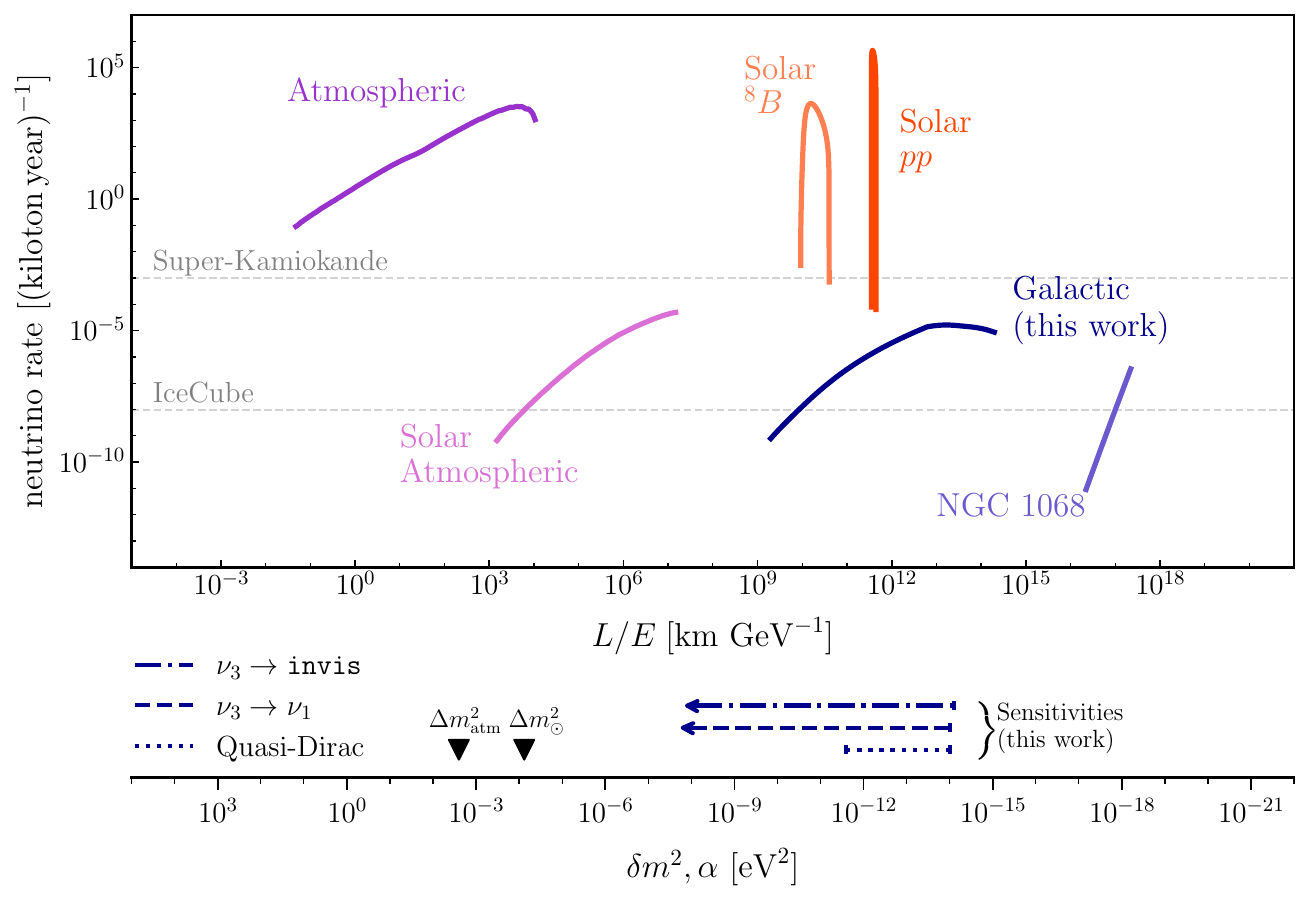}
\caption{\label{fig:LoEplot} \textit{Top panel:} Neutrino rate from various astrophysical sources as a function of $L/E$, in units of $(\mathrm{kiloton~year})^{-1}$. IceCube and Super-Kamiokande exposures are shown as grey dashed lines. Note that the solar atmospheric neutrino flux is the only source that has not been detected yet. The Galactic rate, considered in this work, is plotted in dark blue and is calculated using the \texttt{TANDEM} model (model specifics are included in the main text). The atmospheric, solar atmospheric, and NGC 1068 rates were calculated from fluxes in \cite{Honda:2015anf, Arguelles:2017eao, IceCube:2022der} respectively. The solar rates were calculated from fluxes in \cite{Vitagliano:2019yzm}. \textit{Bottom axis:} We plot the corresponding $\delta m^2$ (QD mass splitting) and $\alpha$ (decay parameter) values that correspond to the given $L/E$ value in the top panel. We plot the solar and atmospheric mass splittings as black triangles. We plot our projected $1\sigma$ sensitivity regions in dark blue for QD (dotted line), visible decay (dashed line), and invisible decay (dot-dashed line) scenarios.}
\end{figure}

In 2023, the IceCube Neutrino Observatory discovered neutrino emission from the Galactic plane \cite{IceCube:2023ame}. With energies $\mathcal{O}(1-10)~\mathrm{TeV}$ and baselines $\mathcal{O}(1-10)~\mathrm{kpc}$, the characteristic $L/E$ profile of Galactic neutrinos is unique among detected neutrino sources, as shown in \cref{fig:LoEplot}. This opens up the possibility of probing long-baseline effects on neutrino propagation that only show up at these $L/E$ scales.

In this work, we show that neutrino telescopes, like IceCube and KM3NeT, can probe novel parameter space in the QD neutrino model and neutrino decay models using the Galactic neutrino flux.

\section{Models}

The diffuse Galactic neutrino emission is produced along a range of baselines, requiring a three-dimensional model of Galactic neutrinos to most accurately simualte the effect of propagation-altering phenomena. The emission templates used in the 2023 discovery only give energy and spatial information, with no information about the neutrino distribution along each line of sight. We use an upcoming three-dimensional model of the Galaxy in neutrinos, called \texttt{TANDEM} (Template of Astrophysical Neutrinos in Distance and Energy in the Milky (Way)) \cite{Carloni:2024aps}, which uses the \texttt{CRPropa} software \cite{AlvesBatista:2022vem} to simulate cosmic ray (CR) production and diffusion in the Galaxy. The model generates a 3D map of the neutrino emissivity, which quantifies the neutrino production rate when the CRs interact with interstellar gas. Integrating the \texttt{TANDEM} emissivity against known Galactic gas models produces a 3D map of the Milky Way in neutrinos. The following figures \cref{fig:binnedeventrates} and \cref{fig:sensitivityplot} were calculated using the \texttt{base} magnetic field model from \cite{Unger:2023cmf} and the HI and HII gas models from \cite{Mertsch:2021bit, Mertsch:2023bit}.

We then account for the propagation-altering effects caused by QD neutrinos and neutrino decays. The probability of a neutrino of flavor $\alpha$ produced a distance $L$ away from Earth with energy $E_\nu$ to be in the $\beta$ flavor state at Earth is given by
\begin{equation}
    P_{\alpha\beta}^\mathrm{QD} = \sum_{i=1}^3 |U_{\alpha i}|^2 |U_{\beta i}|^2 \cos^2 \left( \frac{\delta m_i^2 L}{4 E_\nu}\right), \quad P_{\alpha\beta}^\mathrm{decay} = \sum_{i=1}^3 |U_{\alpha i}|^2 |U_{\beta i}|^2\exp\left\{ - \frac{\alpha_iL}{E_\nu}  \right\}.
\end{equation}
Here, $P^\mathrm{QD}$ denotes the QD scenario, where $\delta m^2$ is the ultrafine active-sterile mass splitting. $P^\mathrm{decay}$ denotes the (invisible) decay scenario, where $\alpha_i$ is the \textit{decay parameter} $\alpha_i \equiv m_i/\tau_i$, with $m_i$ being the $i$th neutrino mass and $\tau_i$ being the $i$th neutrino rest-frame lifetime. The index $i$ runs over the three active neutrino mass states. $U$ is the PMNS matrix. We also consider a case of visible neutrino decay, where further assumptions have to be made on the neutrino mass configuration––namely, we assume that the neutrino masses are \textit{quasi-degenerate}, which occurs when the absolute mass scale is relatively large, on the upper end of the parameter space allowed by KATRIN \cite{KATRIN:2024cdt}. We also assume that neutrino masses follow normal ordering for all scenarios.

We comment on the phenomenology afforded by QD oscillations and neutrino decays on the Galactic neutrino flux, shown in \cref{fig:binnedeventrates}. QD oscillations will lead to visible oscillations in the energy spectrum when $\delta m^2 \sim 4E_\nu/L$––for larger energies, the oscillation probabilities become the constant SM expectation, and for smaller energies, rapid oscillations average out and result in a shift in normalization of $1/2$. Decays will result in attenuations to the $i$th mass state flux when $\alpha_i \gtrsim E_\nu / L$, and in visible decay scenarios a resulting boost to the decay product mass state flux in the same range. In both cases, for the fixed energy ranges and baselines of the Galactic flux, we expect modifications to the signature from such BSM effects to begin when $\delta m^2, \alpha \gtrsim 10^{-15}~\mathrm{eV^2}$. 

We consider a QD scenario where all three active-sterile mass splittings are the same. As a result, we do not expect any modifications to the flavor ratio. However, the decay scenario where all three mass state lifetimes are equal is heavily disfavored \cite{Ivanez-Ballesteros:2023lqa}, so we only consider scenarios where $\nu_3$ is unstable and decays to $\nu_1$. In these scenarios, we expect modifications to the flavor ratio due to decay effects.

\section{Analysis and Results}

We simulate the detector response of two neutrino telescopes, IceCube and KM3NeT/ARCA (henceforth referred to as KM3NeT), to the Galactic neutrino flux under SM, QD, and decay scenarios. Both telescopes detect neutrinos through their emitted Cherenkov light, seeing two main event topologies; tracks, caused by charged current (CC) $\nu_\mu$ interactions, and cascades, caused by CC $\nu_e$ and $\nu_\tau$ interactions, as well as neutral current (NC) interactions of all flavors. IceCube has already seen the GP in a cascade event selection, which has good energy resolution but poor angular resolution \cite{IceCube:2023ame}. It has not been able to detect the GP in a track event selection yet due to the position of the inner GP in the Southern sky, where high atmospheric neutrino backgrounds obscure tracks much more than cascades. KM3NeT will be sensitive to the GP in tracks due to its location in the Northern Hemisphere, placing the inner GP below the horizon and drastically reducing atmospheric neutrino backgrounds. Due to the long lever arm for track events, it will see the GP with sub~degree angular resolution, but poor energy resolution.

\begin{figure}
\centering 
\includegraphics[width=\linewidth]{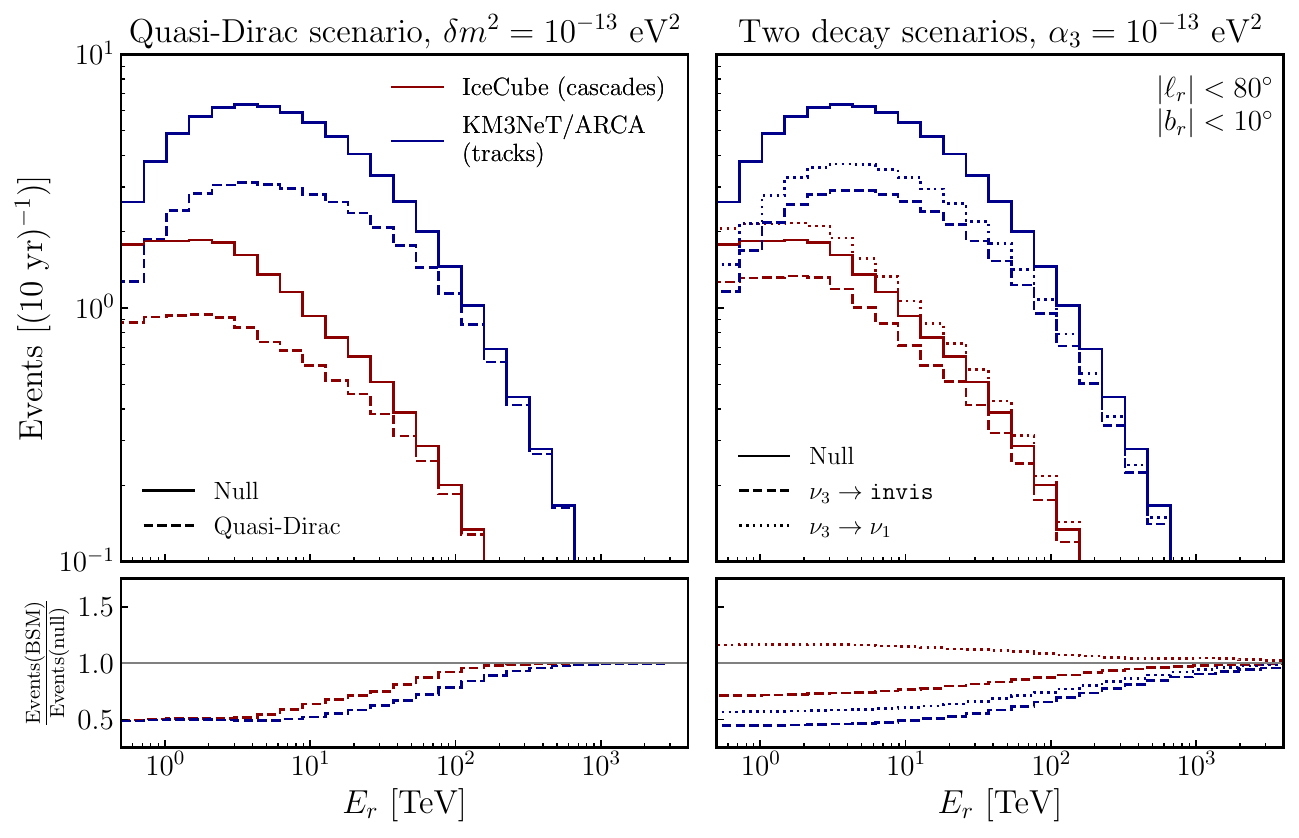}
\caption{\label{fig:binnedeventrates} \textit{Left:} The top panel shows expected event counts binned in reconstructed neutrino energy $E_r$ for 10 years of exposure at IceCube (red) and KM3NeT/ARCA (blue) after convolving with energy and angular resolution. Events are shown for sky window $|\ell_r| < 80^\circ,~|b_r|<10^\circ$. Solid lines show the SM expectation, while dashed lines show the QD expectation for $\delta m^2 = 10^{-13}~\mathrm{eV^2}$. The bottom panel shows the ratio of expected QD events to expected SM events in each energy bin. \textit{Right:} Same as left, but for visible (dotted lines) and invisible (dashed lines) decay scenarios, with $\alpha = 10^{-13}~\mathrm{eV^2}$.}
\end{figure}

High energy resolution is crucial for gaining sensitivity to QD and decay effects to discriminate their respective phenomenology on the neutrino energy spectrum. High angular resolution is important to better discriminate individual baselines, leading to more distinct $L/E$ features in the energy spectrum. A combination of track and cascade samples will afford a flavor ratio component to the analysis, which increases the sensitivity to flavor-altering decay effects. The complementary strengths of IceCube and KM3NeT in these regards motivate an exploration of their combined sensitivity.

\begin{figure}
\centering 
\includegraphics[width=\linewidth]{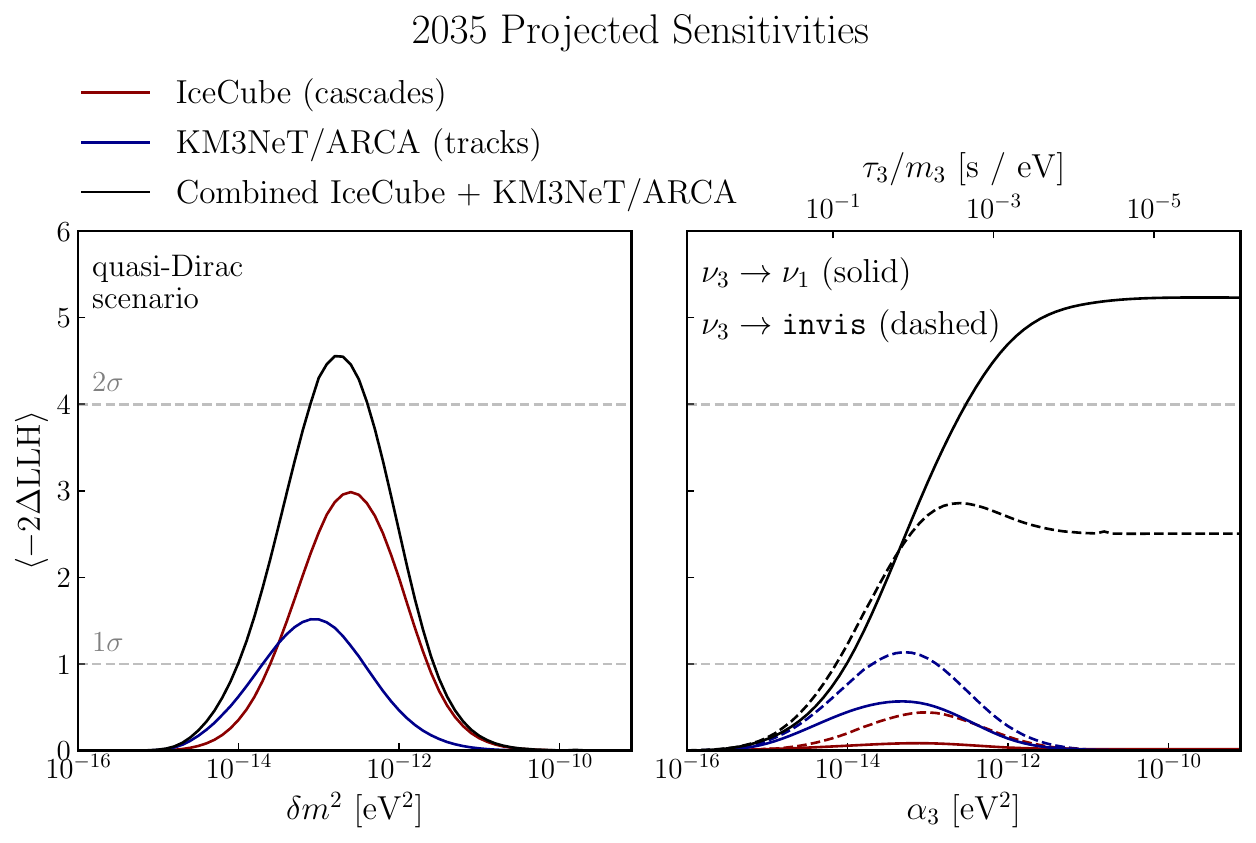}
\caption{\label{fig:sensitivityplot} Sensitivities of IceCube, KM3NeT, and both experiments combined to QD neutrinos and neutrino decay. \textit{Left:} Median test statistic for IceCube cascades (red), KM3NeT tracks (blue), and the combined analysis (black) under a QD scenario. \textit{Right:} Same as left, but for invisible decay of $\nu_3 \to \mathtt{invis}$ (dashed lines) and visible decay of $\nu_3 \to \nu_1$ (solid lines), assuming a quasi-degenerate mass configuration. All sensitivities assume normal ordering.
}
\end{figure}

We model the IceCube detector response using the effective areas and angular and energy resolutions from \cite{IceCube:2023ame}, and we model the full KM3NeT detector following the results of \cite{KM3NeT:2024paj}. To estimate the effect of backgrounds on the eventual sensitivity, we use the H3a\_SYBYLL23C model in \texttt{nuflux} and implement the muon self-veto effect using the tabulated values in \cite{Arguelles:2018awr}. We assume a $(1:2:0)$ $\nu_e:\nu_\mu:\nu_\tau$ flavor ratio at production. 

We note that the current best-fit Galactic neutrino fluxes to similar emission templates like CRINGE \cite{Schwefer:2022zly} predict a total emission $\sim$5 times larger than the emission predicted in \texttt{TANDEM}. The tension in predicted and measured emission might be due to mismodeling of Galactic CR contribution, contribution from unresolved neutrino sources, or a combination of both factors. We compensate for these discrepancies by multiplying our \texttt{TANDEM} flux predictions by a factor of 5. 

We simulate the expected event counts, binned in energy and angle, for IceCube and KM3NeT for a forecast analysis date of 2035, which constitutes 23 years of IceCube data and 8 years of KM3NeT data. Example binned event rates are shown in \cref{fig:binnedeventrates}. We perform a binned likelihood ratio test assuming Poisson statistics and a SM signal. Because of our poor knowledge of the overall Galactic flux normalization, we marginalize over this quantity in our analysis.

We present our preliminary results in \cref{fig:sensitivityplot}, showing both individual sensitivities and the combined IceCube+KM3NeT sensitivity to a 2035 analysis. We consider the QD scenario, an invisible $\nu_3 \to \mathtt{invis}$ decay scenario, and a visible $\nu_3 \to \nu_1$ decay scenario. In the QD scenario, we find that IceCube and KM3NeT, when combined, are sensitive to several decades of unprobed parameter space at $> 1\sigma$ significance, and over one decade at $> 2\sigma$ significance. The forecast decay sensitivity regions reach over $1\sigma$ exclusion sensitivity for decay parameter values of $\alpha \gtrsim 10^{-14}~\mathrm{eV^2}$.

The individual experiment sensitivities add coherently in the QD case, because we considered a scenario that did not induce any flavor-changing phenomena. We lose sensitivity to larger $\delta m^2$ values because in this regime, QD oscillations are all averaged out, yielding a normalization factor difference that is accounted for in our marginalization procedure. In the decay scenarios, the combined sensitivities are greater than the sum of the individual sensitivities. This reflects the increased power of the combined analysis to flavor-changing effects. In the large $\alpha$ regime, we retain sensitivity in a combined analysis because the respective track and cascade best-fit normalizations are different due to their different $\nu_3$ and $\nu_1$ compositions.

\section{Conclusions}

The discovery of the Galactic plane in neutrinos, with its unique baseline and energy profile, paves the way for probing BSM neutrino physics in unexplored parameter space. Here, we show that a global network of neutrino telescopes is sensitive to unexplored parameter space in QD neutrino models, as well as neutrino decay scenarios. The combination of multiple experiments not only increases statistics but also allows for flavor discrimination, which is crucial in breaking normalization degeneracies in flavor-dependent BSM effects. Other planned high-energy neutrino telescopes, like IceCube-Gen2 \cite{IceCube-Gen2:2020qha}, GRAND \cite{GRAND:2018iaj}, and TAMBO \cite{Thompson:2023pnl} will only improve the sensitivity to these novel neutrino models by analyzing the Galactic emission more precisely as well as detecting new neutrino sources with different $L/E$ profiles.

\bibliographystyle{unsrt}
\bibliography{references}

\end{document}